\documentclass{IEEEtran}
\usepackage{graphicx}
\usepackage{cite}
\usepackage{hyperref}
\usepackage{braket}
\usepackage{amssymb}
\usepackage{lipsum}
\usepackage{xcolor}
\usepackage{caption}

\hypersetup{
  colorlinks = true,
  urlcolor   = blue,
  linkcolor  = blue,
  citecolor  = magenta
}

\title{High-fidelity superconducting quantum processors via laser-annealing of transmon qubits}

\author{Eric~J.~Zhang, Srikanth~Srinivasan, Neereja~Sundaresan, Daniela~F.~Bogorin, Yves~Martin, Jared~B.~Hertzberg, John~Timmerwilke, Emily~J.~Pritchett, Jeng-Bang~Yau, Cindy~Wang, William~Landers, Eric~P.~Lewandowski, Adinath~Narasgond, Sami~Rosenblatt, George~A.~Keefe, Isaac~Lauer, Mary~Beth~Rothwell, Douglas~T.~McClure, Oliver~E.~Dial, Jason~S.~Orcutt, Markus~Brink, Jerry~M.~Chow

\vspace{1.5mm} \emph{IBM Quantum, IBM T. J. Watson Research Center, Yorktown Heights, NY 10598, USA}

\vspace{0.0mm} (Dated: December 15, 2020)}

\IEEEaftertitletext{\vspace{-1\baselineskip}}
\date{\today}
\begin{document}

\maketitle
\begin{abstract}
Scaling the number of qubits while maintaining high-fidelity quantum gates remains a key challenge for quantum computing. Presently, superconducting quantum processors with $>$~50-qubits are actively available. For such systems, fixed-frequency transmons are attractive due to their long coherence and noise immunity. However, scaling fixed-frequency architectures proves challenging due to precise relative frequency requirements. Here we employ laser annealing to selectively tune transmon qubits into desired frequency patterns. Statistics over hundreds of annealed qubits demonstrate an empirical tuning precision of 18.5~MHz, with no measurable impact on qubit coherence. We quantify gate error statistics on a tuned 65-qubit processor, with median two-qubit gate fidelity of 98.7\%. Baseline tuning statistics yield a frequency-equivalent resistance precision of 4.7~MHz, sufficient for high-yield scaling beyond 10\textsuperscript{3} qubit levels. Moving forward, we anticipate selective laser annealing to play a central role in scaling fixed-frequency architectures.
\end{abstract}
\section{Introduction}
\label{Introduction}
Recent technological advances have enabled rapid scaling of both the physical number of qubits and computational capabilities of quantum computers \cite{Divincenzo2000,Gambetta2017,DevoretReview2013,QV64_2020}. The distinction between classical and quantum computing arises from the exponentially larger computational subspace available to qubits (quantum bits), which may be set to non-classical superpositions and entangled states. Existing multi-qubit systems built on superconducting circuit quantum electrodynamics (cQED) architectures~\cite{Koch2007, Blais2004} have been utilized in applications ranging from early implementations of Shor's factoring algorithm~\cite{Lucero2012}, to quantum chemistry simulations~\cite{Arguello2019,McArdle2020,Moll2018} and accelerated feature mapping in machine-learning~\cite{Schuld2019,Havlicek2018}. Presently, solid-state cQED-based processors have significantly increased in scale~\cite{Preskill2018,Rosenberg2017}, with dozens of physical qubits demonstrated on a single quantum chip. As gate fidelities improve and eventually reach thresholds required for fault-tolerance, quantum advantage will be exploited to simulate complex molecular dynamics and implement quantum algorithms on practical scales \cite{Shor1999,Grover1996,Karalekas2020}. To track the continual progression of quantum processing power, the \emph{quantum volume} (QV) metric is used as an overall measure of the computational space available for a given processor~\cite{QV64_2020}.

Amongst the major classes of superconducting qubits, fixed frequency transmons~\cite{Koch2007, Gambetta2017} operating in the \emph{E\textsubscript{J}~$\gg$~E\textsubscript{C}} regime are attractive for their low charge dispersion, yielding relatively noise-immune qubits with coherence times (\emph{T\textsubscript{1}},~\emph{T\textsubscript{2}}) exceeding 100~$\mu$s~\cite{Place2020}. Transmon qubits are amenable to high fidelity ($>$~99.9\%) single-qubit gate operations~\cite{Gambetta2017}, and two-qubit entangling cross-resonance (CR) gates~\cite{Tripathi2019} are realized via static qubit-qubit coupling activated by an all-microwave drive scheme~\cite{Gambetta2017,Rigetti2010}. A key requirement to enable high-fidelity CR gates involves the selective addressability of fixed-frequency transmons, as well as precisely engineering their computational $\ket{0}$$\rightarrow$$\ket{1}$ transitions (\emph{f\textsubscript{01}}) for optimal two-qubit interaction~\cite{Gambetta2017}. For example, sub-optimal \emph{f\textsubscript{01}} separation between neighboring qubits reduces ZX coupling strength whilst higher-order static ZZ interactions cause accumulation of two-qubit errors as well as `spectator' error propagation across the lattice~\cite{Sundaresan2020}. The most likely frequency collisions and corresponding tolerance bounds resulting from frequency crowding of lattice transmons have been quantitatively enumerated in~\cite{Hertzberg2020}.

The principal challenge for scaling fixed-frequency architectures is mitigating errors arising from lattice frequency collisions. Typical fabrication tolerances for transmon frequencies range from 1-2\%, with uncertainties dominated by the 2-4\% variation in tunnel junction resistance~\emph{R\textsubscript{n}}~\cite{Hertzberg2020,Kreikebaum2020}. Thermal annealing methods to adjust and stabilize post-fabrication \emph{R\textsubscript{n}} (and correspondingly, transmon frequencies \emph{f\textsubscript{01}}) have been explored previously in~\cite{Muthusubramanian2019,Lehnert1992,Koppinen2007,Orcutt2019}. More recently, the LASIQ (\emph{\textbf{L}aser \textbf{A}nnealing of \textbf{S}tochastically \textbf{I}mpaired \textbf{Q}ubits}) technique was introduced to increase collision-free yield of transmon lattices by selectively trimming (i.e. \emph{tuning}) individual qubit frequencies via laser thermal annealing~\cite{Hertzberg2020}. The LASIQ process sets \emph{R\textsubscript{n}} with high-precision, and \emph{f\textsubscript{01}} could be predicted from \emph{R\textsubscript{n}} according to a power-law relationship resulting from the Ambegaokar-Baratoff relations and transmon theory~\cite{Ambegaokar1963,Koch2007}. An empirical \emph{f\textsubscript{01}(R\textsubscript{n})} scatter of \emph{$\sigma\textsubscript{f}$}~$\simeq$~14~MHz limited the ability to predict \emph{f\textsubscript{01}} from \emph{R\textsubscript{n}}, resulting in a post-tune fequency precision of equivalently $\sim$14~MHz~\cite{Hertzberg2020}.

Here, we demonstrate LASIQ as a scalable process tool used to reduce two-qubit gate errors by systematically trimming lattice transmon frequencies to desired patterns. We show laser tuning results on statistical aggregates of $>$~300 tuned qubits, and demonstrate LASIQ baseline frequency-equivalent resistance tuning precision of 4.7~MHz from empirical \emph{f\textsubscript{01}(R\textsubscript{n})} correlations, reaching this precision with 89.5\% success rate. Based on cryogenic \emph{f\textsubscript{01}} measurements from our laser-tuned processors, we empirically find a frequency assignment precision of 18.5~MHz, which in addition to the LASIQ tuning precision includes all deviations arising from pre-cooldown steps including post-tuned chip cleaning and bonding processes. Our results indicate the precision of trimming \emph{f\textsubscript{01}} is dominated by the residual \emph{f\textsubscript{01}(R\textsubscript{n})} scatter of untuned qubits (\emph{$\sigma\textsubscript{f}$}~=~18.1~MHz), and is not limited by the LASIQ trimming process itself.

In addition to scaling the number of tuned qubits, we measure functional parameters of multi-qubit chips (coherence and two-qubit gate-fidelity) to ensure high processor performance. We assess the impact of LASIQ tuning on qubit coherence using a collection of composite (partially tuned) processors, showing aggregate mean \emph{T\textsubscript{1}} and \emph{T\textsubscript{2}} times of 79~$\pm$~16~$\mu$s and 69~$\pm$~26~$\mu$s respectively, with no statistically significant variation between the tuned and untuned groups. Our LASIQ process has been broadly utilized for precise frequency control of post-fabricated 27-qubit \emph{Falcon} processors, including the recent QV-128 \emph{ibmq\_montreal} system~\cite{QV64_2020,QV128_unpublished}. We demonstrate LASIQ scaling capabilities by tuning a 65-qubit \emph{Hummingbird} processor (cloud-accessible as \emph{ibmq\_manhattan}), with a median two-qubit gate fidelity of 98.7\% based on randomized benchmarking \cite{Knill2008,Gambetta2017}. As a scalable frequency trimming tool for fixed frequency transmon architectures, we envision the LASIQ process to be widely implemented in future generations of superconducting quantum systems.

\begin{figure}[t!]
\label{fig:1}
\centering
\includegraphics[width=3.75in, trim= {0.1in 0.1in 8.65in 0.06in}, clip]{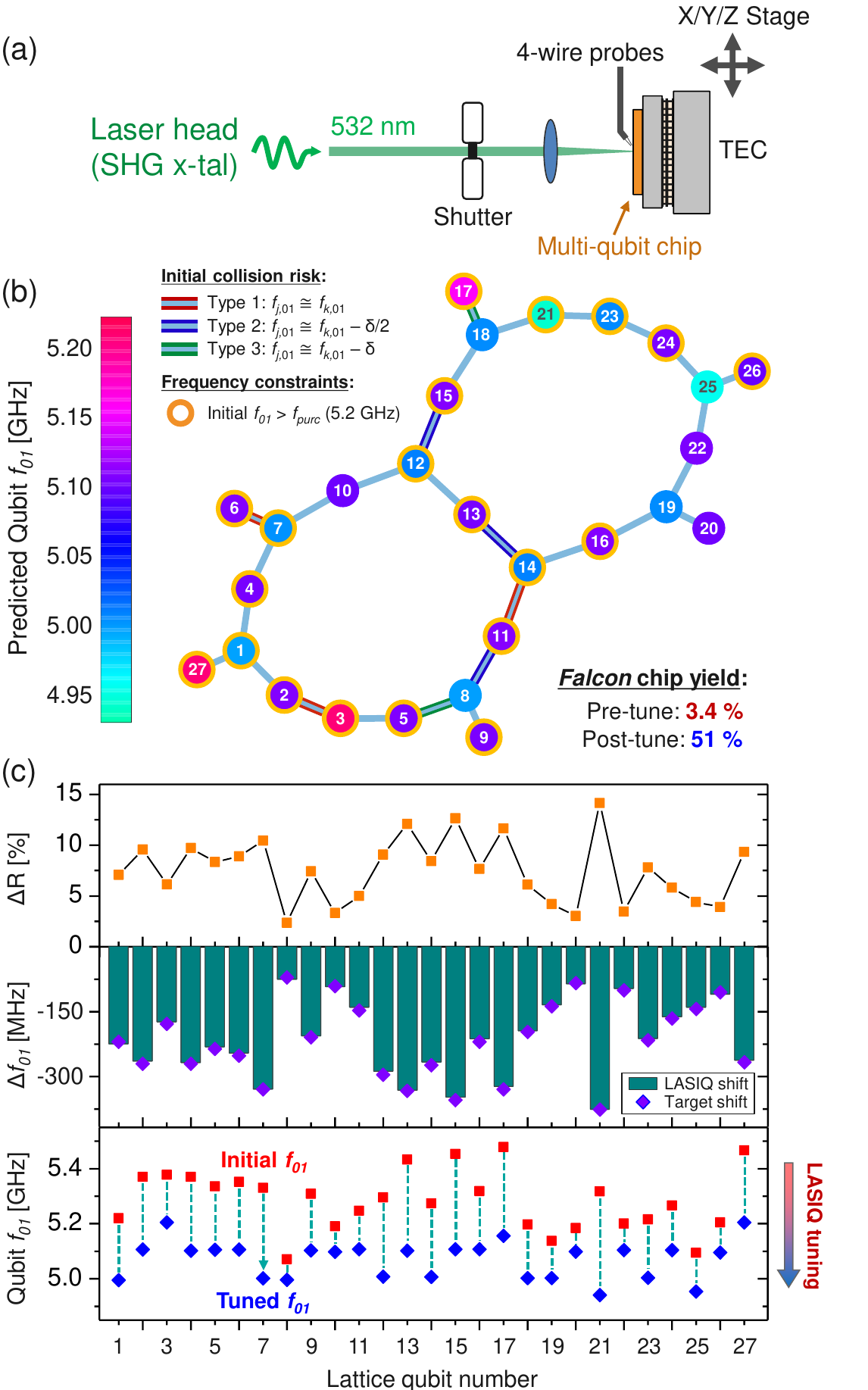}
\caption{Example of a LASIQ anneal process. (a)~Outline of the laser trimming setup~\cite{Hertzberg2020}. A 532 nm second-harmonic generation laser is sequentially focused on the junctions of a multi-qubit quantum processor, with thermal annealing to selectively decrease qubit frequencies (\emph{f\textsubscript{01}}) for collision avoidance. (b)~Example of a tuned 27-qubit \emph{Falcon} lattice. Final predicted \emph{f\textsubscript{01}} are depicted as a heatmap, with initial high-risk NN collision pairs highlighted, and orange outlines indicating initial \emph{f\textsubscript{01}} above the bandwidth of Purcell protection. Post-LASIQ, collision and frequency constraints are resolved. (c)~Detail of qubit anneals. The bottom panel indicates the initial (red) and final (blue) predicted \emph{f\textsubscript{01}} showing the qubits tuned to distinct frequency setpoints. The middle panel indicates the tuning distance (monotonic negative shifts), along with the desired target shifts (purple diamonds), with an RMS deviation (i.e. frequency-equivalent resistance tuning precision) of 4.8~MHz, as determined from empirical \emph{f\textsubscript{01}(R\textsubscript{n})} correlations. The top panel depicts the corresponding junction resistance shifts, achieving tuning ranges up to 14.2\%.}
\end{figure}

\section{Results}
\label{Results}

\subsection{Tuning a 27-qubit Falcon processor}
\label{Sec II-A}
As a practical demonstration of frequency trimming, a 27-qubit \emph{Falcon} processor is tuned to predicted frequency targets using the LASIQ setup shown in \hyperref[fig:1]{Fig.~1(a)} (described in \hyperref[meth:LaserAnnealing]{Methods}). The \emph{Falcon} chip series are based on a heavy-hexagonal lattice, which contains the distance-3 hybrid surface and Bacon-Shor code for error correction~\cite{Chamberland2020}. For this demonstration, all measurements were performed at ambient conditions, with resulting tuned frequencies estimated from empirical \emph{f\textsubscript{01}(R\textsubscript{n})} correlations. The tuned lattice is depicted in \hyperref[fig:1]{Fig.~1(b)}, with a color heatmap indicating the post-LASIQ frequency predictions. Nearest-neighbor (NN) qubit frequency spacing ($\Delta$\emph{f\textsubscript{NN}}) can visually be seen to be separated by 50~MHz~$\leq$~$|$$\Delta\emph{f\textsubscript{NN}}$$|$~$\leq$~250~MHz, placing qubits comfortably within the straddling regime for high-ZX interaction \cite{Magesan2020}. NN collisions have been avoided with twice the collision tolerance (2$\Delta\textsubscript{c}$, see \hyperref[sec:supplement]{Supplementary Information}) from bounds described in~\cite{Hertzberg2020}, to protect against two-qubit state hybridization and improve chip yield. Prior to tuning, high-risk pairs within 2$\Delta\textsubscript{c}$ collision bounds have edge borders highlighted as indicated in the lattice graph, and are resolved after the LASIQ process is complete. Based on Monte Carlo models with a conservative post-tune spread estimate of \emph{$\sigma\textsubscript{f}$}~=~$20$~MHz (\hyperref[Sec. II-B]{Sec.~II-B}), we demonstrate substantial increase in the yield rate from 3.4\% to 51\% for achieving zero NN collisions, corresponding to 15$\times$ yield improvement after LASIQ tuning (see \hyperref[sec:supplement]{Supplementary Information}). The target qubit \emph{f\textsubscript{01}} patterns corresponding to \hyperref[fig:1]{Fig.~1(b)} are shown in the bottom panel of \hyperref[fig:1]{Fig.~1(c)}, where initial \emph{f\textsubscript{01}} (red) are progressively tuned until they reach pre-defined and distinct frequency setpoints (blue).

In addition to NN collision avoidance, all qubits have been tuned to targets at or below $\emph{f\textsubscript{purc}}$~=~5.2~GHz, which in our \emph{Falcon} design is the cutoff for maintaining good Purcell protection and avoiding radiative qubit relaxation~\cite{Jeffrey2014}. Due to the monotonic increase of junction resistance (\emph{R\textsubscript{n}}) intrinsic to the laser anneal process, qubit frequencies are `trimmed' (i.e. reduced) to desired \emph{f\textsubscript{01}} values. The LASIQ process is engineered to proceed until \emph{R\textsubscript{n}} for all junction reside within 0.3\% of the target \emph{R\textsubscript{T}} (corresponding to $\sim$10~MHz for typical \emph{f\textsubscript{01}(R\textsubscript{n})} correlation coefficients), although a nominal LASIQ approach will typically outperform this upper precision bound \hyperref[Sec. II-B]{(Sec.~II-B)}. The center panel of \hyperref[fig:1]{Fig.~1(c)} shows desired target shifts (purple diamonds) superimposed on final predicted frequency tuning amplitudes (green bars), indicating excellent agreement with desired target values. The root-mean square (RMS) \emph{R\textsubscript{n}} deviation from \emph{R\textsubscript{T}} for this processor is $0.16$\%, corresponding to a frequency-equivalent resistance tuning precision of 4.8~MHz (using nominal \emph{f\textsubscript{01}(R\textsubscript{n})} power-law coefficients for this sample), and is consistent with LASIQ baseline tuning precision as described in \hyperref[fig:2]{Fig.~2}. As seen from the center panel, target \emph{f\textsubscript{01}} shifts range from 75~MHz (Qubit~8) to 375~MHz (Qubit~21) in \hyperref[fig:1]{Fig.~1(c)}, corresponding to resistance shifts up to \emph{$\Delta$R\textsubscript{n}}~=~14.2\%. Based on separate calibration measurements over standard junction arrays, resistance tuning peaks at $\sim$14\%, and therefore tuning plans for each processor are designed to be constrained within attainable targets. Upon completion of the LASIQ process, a typical quantum processor is cooled and screened for coherence and single/two-qubit gate fidelity, and assessments of quantum volume are performed \cite{QV64_2020}. Two-qubit gate error statistics of a tuned and operational 65-qubit \emph{Hummingbird} processor are shown in \hyperref[Sec. II-C]{Sec.~II-C}.

\subsection{LASIQ tuning precision}
\label{Sec. II-B}
\begin{figure}[t!]
\label{fig:2}
\centering
\includegraphics[width=7.7in, trim= {0.22in 0.22in 3.8in 0.25in}, clip]{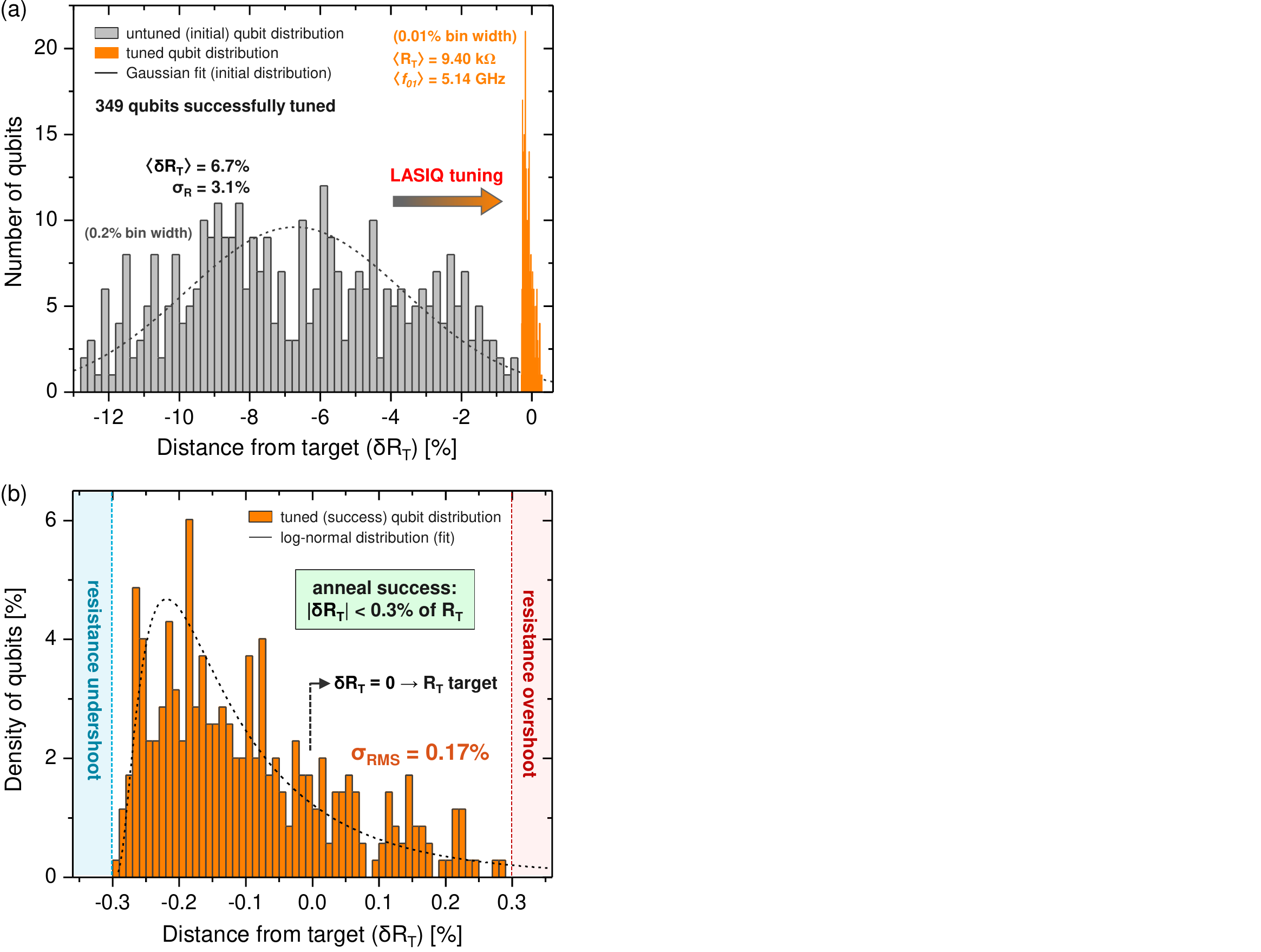}
\caption{LASIQ tuning outcome statistics. (a) The initial distribution (gray) of qubits that were successfully tuned to target (orange). The distance from target \emph{$\delta$R\textsubscript{T}} is the tuning differential normalized to the final target resistance \emph{R\textsubscript{T}}. Orange bars indicate the final distribution (20$\times$ reduced bin width for clarity) and shows the 349 qubits tuned to success. (b) Expanded view of the orange distribution shown in (a). Anneal success is defined as a tuned resistance within 0.3\% of \emph{R\textsubscript{T}}, which was reached by all displayed qubits, and 89.5\% of the 390 tuned qubits (details in \hyperref[sec:supplement]{Supplementary Information}). The blue/red regions indicate undershoot/overshoot respectively. A log-normal fit is shown by the black curve, which supports the interpretation of LASIQ tuning as an incremental resistance growth process.}
\end{figure}

In prior work, tuning precision estimates were limited by the imprecision in predicting \emph{f\textsubscript{01}} from room-temperature \emph{R\textsubscript{n}} measurements~\cite{Hertzberg2020}. Thus, only an upper bound of tuning precision could be determined (\emph{$\sigma\textsubscript{f}$}~$\simeq$~14~MHz). Presently, we address the limits of LASIQ tuning precision as limited by the process itself, rather than our ability to predict cryogenic \emph{f\textsubscript{01}(R\textsubscript{n})}. We analyze a large sample of 390 tuned qubits, 349 of which successfully tuned to target (defined in \hyperref[Sec II-A]{Sec. II-A} as achieving junction resistance within 0.3\% of \emph{R\textsubscript{T}}), yielding an aggregate tuning success rate of 89.5\% for this experiment. The initial and final distributions of successfully tuned qubits are shown in \hyperref[fig:2]{Fig.~2}. We note here that a full range of tuning (\emph{$\Delta$R} up to $\sim$14\% with respect to initial junction \emph{R\textsubscript{n}}) was performed for this experiment, with anticipated failure rates being weighted towards tuning extremes (\emph{$\Delta$R}~$>$~14\% increases undershoot risk, whilst low tuning \emph{$\Delta$R}~$<$~1\% results in increased overshoot risk). Such extremes are readily avoided during \emph{f\textsubscript{01}} tuning plan assignment. A detailed analysis of tuning statistics and success rates has been performed for all qubits (see \hyperref[sec:supplement]{Supplementary Information}).

\begin{figure}[t!]
\label{fig:3}
\centering
\includegraphics[width=6.8in, trim= {0.17in 1in 3.8in 0.85in}, clip]{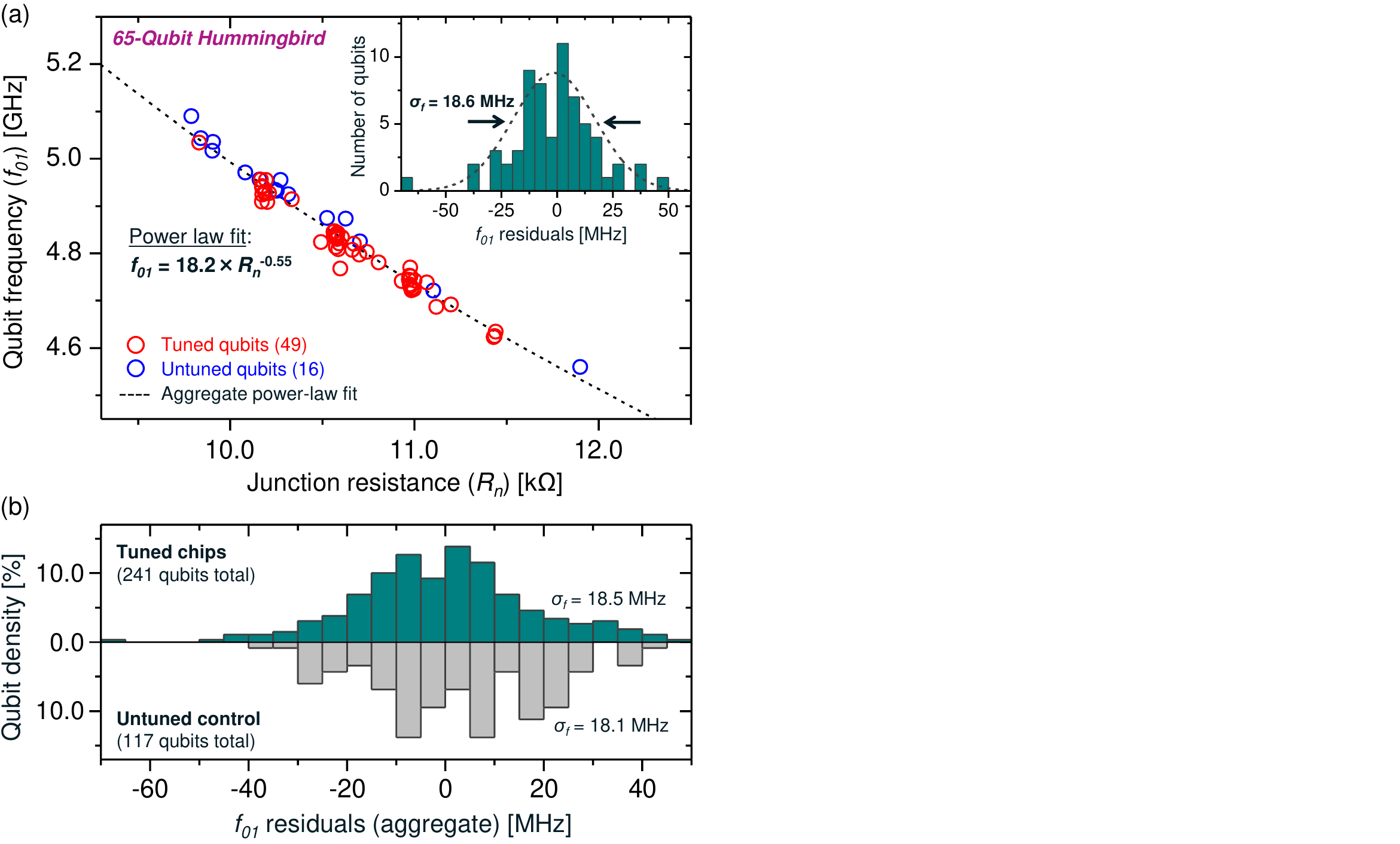}
\caption{Frequency assignment precision based on statistical aggregates of tuned 27-qubit \emph{Falcon} and 65-qubit \emph{Hummingbird} processors. (a) Resistance (\emph{R\textsubscript{n}}) to frequency (\emph{f\textsubscript{01}}) correlation for a tuned \emph{Hummingbird} processor. Cryogenic \emph{f\textsubscript{01}} measurements are plotted against measured junction resistances \emph{R\textsubscript{n}}, with a power-law curve superimposed on the measured data. Both tuned (49 qubits) and untuned (16) qubits are depicted. The inset shows a histogram of residuals with a standard deviation of 18.6~MHz, indicating the practical precision to which we may assign qubit frequencies. (b) The top panel shows statistical precision analysis performed for a total of 241 tuned qubits from a combination of \emph{Falcon} and \emph{Hummingbird} chips, with aggregate \emph{f\textsubscript{01}} residuals from individual power-law regressions for each chip. The bottom panel shows identical analysis performed for 117 untuned qubits from both processor families. Cryogenic \emph{f\textsubscript{01}} measurements yield 18.5~MHz and 18.1~MHz spread for tuned and untuned qubits respectively, indicating that the LASIQ process does not significantly impact the overall spread of qubit frequencies prior to preparatory chip cleaning, bonding and cooldown processes.
}
\end{figure}

To understand the consequences of the incremental approach to \emph{R\textsubscript{T}}, aggregate tuning statistics of the 349 successfully tuned qubits are depicted in \hyperref[fig:2]{Fig. 2(a)}. \emph{R\textsubscript{n}} of the initial qubits (gray histogram) are tuned to a tight tolerance about the targets (orange), as displayed on a \emph{R\textsubscript{T}} normalized scale. A Gaussian fit approximates the initial resistance distribution, yielding a mean fractional tuning distance of 6.7\% (with respect to \emph{R\textsubscript{T}}). The final distribution is magnified in \hyperref[fig:2]{Fig. 2(b)} with a superimposed log-normal curve fit, a characteristic distribution for fractional growth processes consistent with incremental anneals during the LASIQ process. Averaging over the entire distribution yields a RMS error deviation \emph{$\sigma$\textsubscript{R}}~=~0.17\% from \emph{R\textsubscript{T}}, corresponding to \emph{$\sigma$\textsubscript{f}}~=~$\partial f\textsubscript{\emph{01}} / \partial R\textsubscript{\emph{n}} \cdot \emph{$\sigma$\textsubscript{R}}$~=~4.7~MHz as determined by empirical \emph{f\textsubscript{01}(R\textsubscript{n})} correlations from our 65-qubit \emph{Hummingbird} processor (\hyperref[fig:3]{Fig. 3}). Our determination of LASIQ resistance precision shows that the fundamental tuning performance is well below the upper bound of $\sim$14~MHz determined in \cite{Hertzberg2020}, where it was also noted that this bound was dominated by residual scatter in the prediction of cryogenic \emph{f\textsubscript{01}} from room temperature \emph{R\textsubscript{n}}. Based on Monte Carlo simulations, a precision of $\leq$~6~MHz is required for high-yield scaling for quantum processors beyond 10\textsuperscript{3} qubits~\cite{Hertzberg2020}. Therefore, our baseline frequency-equivalent resistance precision of \emph{$\sigma$\textsubscript{f}}~$<$~5~MHz demonstrates LASIQ as a viable post-fabrication trimming process for high-yield scaling of fixed frequency transmon processors.

In addition to the residual \emph{f\textsubscript{01}(R\textsubscript{n})} prediction scatter, a number of preparatory cleaning, bonding and processor mounting steps are undertaken for our quantum processors in the pre-cooldown stage. A natural question therefore arises as to the practical precision achieved for frequency assignment with the inclusion of such processes. \hyperref[fig:3]{Fig. 3(a)} shows cryogenic \emph{f\textsubscript{01}} plotted against post-LASIQ measurements of \emph{R\textsubscript{n}} on a 65-qubit \emph{Hummingbird} processor, which comprised 49 tuned qubits and 16 untuned qubits. Post-tuning, the processor underwent plasma cleaning and flip-chip bump-bonding to an interposer layer prior to mounting in a dilution refrigerator for cooldown and screening (see \hyperref[meth:FabricationAndPreparation]{Methods}). The entire process occurred within a 24-hour span to minimize the impact of aging and drift on the junction resistances (and therefore qubit frequencies). A power-law fit (dashed curve) conforms well to both the tuned and untuned qubits, indicating that no appreciable relative shifts due to LASIQ tuning have occurred, and that the same \emph{f\textsubscript{01}(R\textsubscript{n})} prediction may be adequately utilized in both cases. We note the slight deviation of the power-law fit exponent (-0.55) from the nominal one-half expected from the Ambegaokar-Baratoff relations and transmon theory \cite{Ambegaokar1963,Koch2007}, which is attributed to non-idealities in predicting \emph{f\textsubscript{01}} from room-temperature \emph{R\textsubscript{n}}. Nevertheless, this deviation to first-order does not affect the relative frequency spacing between qubits, and therefore does not impact the precision to which we assign qubit frequency spacing for collision avoidance. The effective \emph{f\textsubscript{01}} assignment precision may be determined by the residuals of the power-law fit as shown in the inset of \hyperref[fig:3]{Fig. 3(a)}, outlined by a Gaussian distribution with \emph{$\sigma$\textsubscript{f}}~=~18.6~MHz spread, corresponding to 0.38\% of the mean qubit frequency (4.84~GHz), and defines the practical \emph{f\textsubscript{01}} assignment precision for this chip.

\hyperref[fig:3]{Fig. 3(b)} shows a similar analysis performed over a statistical sampling of 241 qubits from 7 \emph{Falcon} and 2 \emph{Hummingbird} tuned processors. Each processor sample was fit to an individual \emph{f\textsubscript{01}(R\textsubscript{n})} curve and residuals are aggregated in the upper histogram (dark green), with 1-\emph{$\sigma$\textsubscript{f}} spread of 18.5~MHz, consistent with the single-processor sample observed in \hyperref[fig:3]{Fig. 3(a)}. Results from control (untuned) qubits are shown in the bottom panel histogram (gray), which are \emph{f\textsubscript{01}(R\textsubscript{n})} residuals extracted from 117 untuned qubits on composite (partially tuned) processors, yielding 1-\emph{$\sigma$\textsubscript{f}} spread of 18.1~MHz. We may therefore conclude that the \emph{f\textsubscript{01}} assignment imprecision resulting from our LASIQ process is a negligible contributor to the overall frequency spread of the qubits. We note that although our residual value for both tuned and untuned samples is larger than the $\sim$14~MHz demonstrated in~\cite{Hertzberg2020}, our \emph{Falcon} and \emph{Hummingbird} processors undergo significantly more preparatory steps prior to cooldown and a certain amount of deviation is anticipated. The prediction imprecision of \emph{f\textsubscript{01}(R\textsubscript{n})} remains the dominant contributor to overall spread, with relatively smaller contributions from post-tune drifts and pre-cooldown processes. Significant room remains for improving frequency predictions to reach single-MHz levels dictated by the $\sim$5~MHz LASIQ baseline frequency-equivalent tuning precision as shown in \hyperref[fig:2]{Fig. 2(b)} (details in \hyperref[Discussion]{Sec.~III}).

\subsection{Qubit coherence and gate fidelity}
\label{Sec. II-C}
\begin{figure}[t!]
\label{fig:4}
\centering
\includegraphics[width=7in, trim= {0.25in 3.97in 2.3in 0in}, clip]{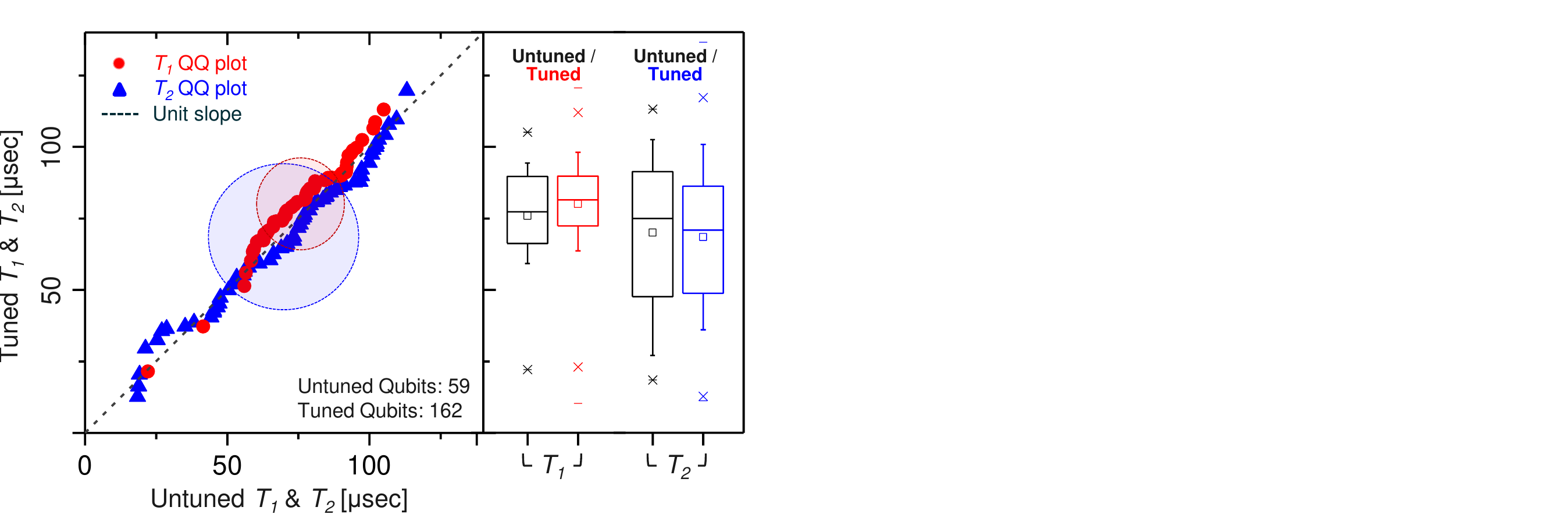}
\caption{Impact of LASIQ tuning on qubit relaxation (\emph{T\textsubscript{1}}, red) and dephasing (\emph{T\textsubscript{2}}, blue), using composite (partially tuned) \emph{Hummingbird} processors. The left panel shows a quantile-quantile (QQ) plot of tuned and untuned qubits. Good linearity with respect to unit slope indicates similar (\emph{T\textsubscript{1}}, \emph{T\textsubscript{2}}) distributions of both tuned and untuned qubits. The shaded ovals are visual indicators of 1-$\sigma$ spread in relaxation and dephasing times. For tuned (untuned) qubits, $\langle$\emph{T\textsubscript{1}}$\rangle$~=~ 80~$\pm$~16~$\mu$s (76~$\pm$~15~$\mu$s) and $\langle$\emph{T\textsubscript{2}}$\rangle$~=~68~$\pm$~25~$\mu$s (70~$\pm$~26~$\mu$s), with robust agreement between mean values within statistical error bounds. The right panel box chart (interquartile box range with 10-90\% whiskers, 1-99\% outliers indicated by crosses and minima/maxima indicated by horizontal delimiters) demonstrates the negligible effect of the LASIQ process on qubit coherence. A total of 59 untuned and 162 tuned qubits from 4 \emph{Hummingbird} chips are displayed, corresponding to a total of 221 measured qubits.}
\end{figure}

Maintaining high qubit coherence is an essential component of high-fidelity single- and two-qubit gates, in addition to precise frequency control. To empirically determine the impact of laser-tuning on qubit coherence, a composite (partially tuned) set of 4 \emph{Hummingbird} processors were cooled and coherence was assessed. Out of a total 221 measured qubits from the 4 composite processors, 162 were LASIQ-tuned and 59 were left untuned (73\% fractional tuning rate). The statistical aggregate is shown in \hyperref[fig:4]{Fig.~4} where tuned and untuned relaxation (\emph{T\textsubscript{1}}, red) and dephasing (\emph{T\textsubscript{2}}, blue) times are correlated on a quantile-quantile plot (left panel). Good correspondence is observed with respect to linear unity slope, indicating that distributions of coherences remain unaffected by tuning. Visual indicators of the mean $\langle$\emph{T\textsubscript{1}}$\rangle$ and $\langle$\emph{T\textsubscript{2}}$\rangle$ and 1-$\sigma$ bounds are shown in the shaded ovals, with aggregate (including both tuned and untuned qubits) relaxation and dephasing times of 79~$\pm$~16~$\mu$s and 69~$\pm$~26~$\mu$s respectively. A direct comparison between tuned and untuned qubits indicates that for tuned (untuned) qubits, $\langle$\emph{T\textsubscript{1}}$\rangle$~=~ 80~$\pm$~16~$\mu$s (76~$\pm$~15~$\mu$s) and $\langle$\emph{T\textsubscript{2}}$\rangle$~=~68~$\pm$~25~$\mu$s (70~$\pm$~26~$\mu$s). Corresponding box-plots (interquartile box range, with 10-90\% whiskers) are shown in the right panel, demonstrating that within statistical error, no variation in qubit coherence is observed due to the LASIQ process.

\begin{figure}[t!]
\label{fig:5}
\centering
\includegraphics[width=7in, trim= {0in 2.2in 1.5in 0.1in}, clip]{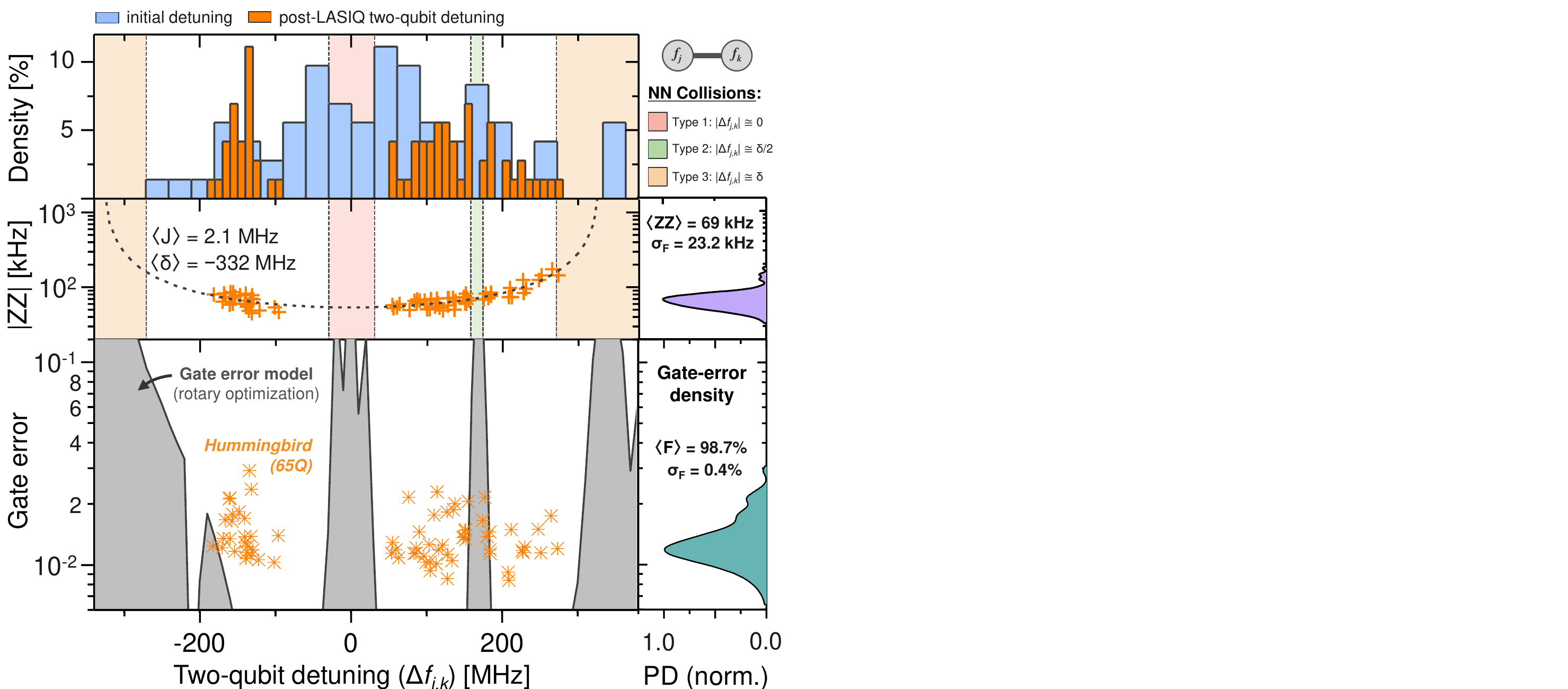}
\caption{Gate errors of a 65-qubit \emph{Hummingbird} processor after LASIQ tuning. The bottom panel shows measured CNOT gate errors as a function of two-qubit detuning (orange points), yielding a median gate fidelity of 98.7\% for the LASIQ-tuned \emph{Hummingbird}. The shaded (gray) regions indicate approximative error-rate projections based on CR gate error modeling \cite{Pritchett2020}, incorporating typical qubit interaction parameters (frequency and anharmonicity, qubit coupling, gate times), with optional rotary echo pulsing for error minimization. The middle panel depicts the achieved ZZ distribution, indicating well-tailored separation near null-detuning (type-1 NN collision), while maintaining a tight ZZ spread with 69~kHz median. The top panel indicates the distribution of tuned two-qubit \emph{f\textsubscript{01}} separation (orange), along with the initial (pre-LASIQ) distribution (blue), indicating high-density of collisions and ZZ amplitudes prior to LASIQ tuning.}
\end{figure}

As a practical demonstration of LASIQ tuning capabilities, a 65-qubit \emph{Hummingbird} processor is laser-tuned and operationally cloud-accessible as \emph{ibmq\_manhattan}. In a similar fashion to that shown in \hyperref[fig:1]{Fig. 1(a)}, the LASIQ tuning plan is generated by ensuring avoidance of nearest-neighbor (NN) level degeneracies whilst maintaining level separation within the straddling regime~\cite{Magesan2020}. Post-LASIQ, the 65-qubit processor was cooled and qubit frequencies were measured, with density of frequency detuning between two-qubit gate pairs shown in the top panel of \hyperref[fig:5]{Fig. 5} (orange, 10~MHz bin width), along with the initial two-qubit detuning (blue, 30~MHz bins). As-tuned, our processor consists of 72 operational two-qubit CR gates, with gate durations ranging from 250 to 600~ns. Collision boundaries (2$\Delta\textsubscript{c}$ bounds as derived from \cite{Hertzberg2020,Magesan2020}, see \hyperref[sec:supplement]{Supplementary Information}) for NN degeneracies are shown in the background, consistent with those utilized in \hyperref[fig:1]{Fig. 1(a)} indicating that $\sim$20\% of gates are within `high-risk' collision zones. Indeed, Monte Carlo yield modeling of the untuned (as-fabricated) 65-qubit \emph{Hummingbird} chip indicates an average of 12 NN collisions (assuming nominal 20~MHz spread with $\Delta\textsubscript{c}$ collision bounds), with an effectively null yield of zero NN collisions. These yield and collision outcomes are typical of untuned processor samples of this size. Post-LASIQ, the collisions are resolved, as evidenced by the high median two-qubit gate fidelity (98.7\%) shown in the lower panel. The corresponding ZZ statistics are shown in the middle panel, indicating good separation near null-detuning (type-1 collision), whilst maintaining tight ZZ distribution with median 69~kHz ($\pm$23.2~kHz). We note that transmon-transmon coupling suffers from static-ZZ (i.e. `always-on') error; however, within the $\Delta$\emph{f\textsubscript{j,k}}~$<$~$\delta$ regime (anharmonicity $\delta$~$\simeq$~\textendash330~MHz, see \hyperref[meth:FabricationAndPreparation]{Methods}), tailoring the ZZ distribution to present magnitudes ($\sim$70~kHz) with collision-free detuning is sufficient to yield gate errors approaching 10\textsuperscript{-2} for typical two-qubit CR gate durations of $\sim$400~ns, indicating that our two-qubit gate-fidelities are near levels set by coherence~\cite{Sundaresan2020}.

Two-qubit gate errors as a function of qubit-pair detuning are displayed in \hyperref[fig:5]{Fig. 5 (bottom panel)}, with their associated distribution shown in the adjacent probability density map (right). Notably, our detuning distribution shows more gates with positive control-target detuning, as expected given the greater ZX interaction in the positive detuning region. A CR gate error model is also depicted in the figure (see \hyperref[meth:GateErrorModel]{Methods}), where the shaded background (gray) indicates low-fidelity regions consistent with known frequency collisions described in~\cite{Hertzberg2020}. As part of our model, an optimizer routine incorporates a standard CR echo sequence with optional target rotary pulsing to determine usable two-qubit detuning regions within the straddling regime. For a gate-error of 1\%, a total usable frequency range of 380~MHz is available (optimized using \emph{J}~=~1.75~MHz in the depicted model), which reduces to 350~MHz (130~MHz) for error targets of 0.5\% (0.1\%). Finally, we note that our depicted model does not incorporate the impacts of classical crosstalk and coherence, the latter of which limits gate error in the desirable detuning regions. We note that although our collisions bounds and unitary gate model have similar qualitative outcomes, further work is required to determine exact collision constraints and identify high-fidelity detuning regimes as lattice sizes are progressively increased.

\section{Discussion}
\label{Discussion}

Significant yield improvement and high two-qubit gate fidelities for both \emph{Falcon} and \emph{Hummingbird} processors demonstrate LASIQ as an effective post-fabrication frequency trimming technique for multi-qubit processors based on fixed-frequency transmon architectures. Selective laser annealing offers a compelling and scalable solution to the problem of frequency crowding, with LASIQ being readily adaptable to the scaling of qubits on progressively larger quantum processors. Based on Monte Carlo models \cite{Hertzberg2020} the 4.7~MHz frequency-equivalent resistance tuning precision of LASIQ allows high-yield scaling up to and beyond the 1000-qubit level. At present, cryogenic \emph{f\textsubscript{{01}}} measurements indicate a practical precision of 18.5~MHz; however, statistical analysis of cryogenic-to-ambient thermal cycling in a dilution refrigerator yields a re-cool stability of 5.7~MHz for our multi-qubit processors, which may be leveraged to obtain \emph{f\textsubscript{01}(R\textsubscript{n})} predictions with similar accuracy and approach the baseline frequency assignment precision allowed by LASIQ. Finally, we note that despite our present emphasis on nearest-neighbor (NN) collisions, errors arising from next-NN qubits (e.g. spectator collisions) are a rising concern with lattice scaling, and future work will incorporate tuning plans to minimize such errors, as well as lattice frequency pattern optimization for yield maximization.

\section{Methods}
\label{sec:IV}

\subsection{Laser annealing for transmon frequency allocation}
\label{meth:LaserAnnealing}

The LASIQ frequency tuning system is depicted in \hyperref[fig:1]{Fig.~1(a)}, whereby a diode-pumped solid-state laser (532~nm) is actively aligned and focused on a multi-qubit quantum processor to selectively anneal individual transmon qubits~\cite{Hertzberg2020}. A timed shutter precisely controls the anneal duration and monotonically increments the junction resistance over multiple exposures, with intermediate resistance measurements to aid the adaptive approach to target resistances (\emph{R\textsubscript{T}}). Diffractive beam shaping is optionally available to aid in the uniform thermal loading of the junction \cite{RosenblattOrcutt2019}. The LASIQ process is fully automated and tunes an entire chip to completion, which occurs when each junction is annealed to within 0.3\% tolerance around \emph{R\textsubscript{T}} (corresponding to $\sim$10~MHz for typical \emph{f\textsubscript{01}(R\textsubscript{n})} correlations).

\subsection{Preparation and screening of multi-qubit processors}
\label{meth:FabricationAndPreparation}

The Josephson junctions are fabricated via standard electron beam lithographic patterning followed by two-angle shadow evaporation with intermediate oxidation for Al/AlO\textsubscript{2}/Al junction formation \cite{Dolan1977}. Typical lateral dimensions of the junction are $\sim$100~nm, which yield as-fabricated frequency deviations near 2\%, corresponding to $\sim$100~MHz of qubit frequency spread. Transmon qubit designs follow those described in~\cite{QV64_2020,Hertzberg2020}, with coupled qubits in a fixed-frequency lattice architecture. Nominal qubit anharmonicities are engineered near $\delta$~$\simeq$~\textendash330~MHz. Following junction evaporation, pre-screening of viable processor candidates are performed via room-temperature resistance measurements of all Josephson junctions, with transmon qubit frequencies estimated through wafer-level resistance-to-frequency power-law fits from prior control experiments. Candidates are ranked based on suitability for LASIQ tuning into collision-free tuning plans and within the Purcell filter bandwidth, all under the constraint of tuning range limits (typical maximum resistance tuning of $\sim$14\%). Following this pre-selection process, a frequency plan is generated and tested through Monte Carlo collision and yield modeling \cite{Hertzberg2020}. Automated LASIQ tuning is performed as defined by the tuning plan for each transmon qubit to a junction resistance precision band of 0.3\%, which yields an as-tuned RMS frequency-equivalent resistance precision $\sim$5~MHz (\hyperref[Sec. II-B]{Sec.~II-B}). Post-LASIQ, the multi-qubit processors are plasma-cleaned and flip-chip bonded to an interposer layer for Purcell filtering and readout. The bonded processors are mounted into a dilution refrigerator for cryogenic screening, including coherence measurements and single- and two-qubit gate-error analysis. The post-LASIQ cleaning, bonding and mounting steps are coordinated to occur within a 24-hour span to minimize junction aging and drifts, thereby maintaining the qubits faithfully outside collision bounds of the intended tuning frequency plans.

\subsection{Gate-error model}
\label{meth:GateErrorModel}

Our gate error estimates are the result of time-domain simulations of the Schrodinger equation within the full Duffing model of two coupled transmon qubits~\cite{Magesan2020}. CR pulses are modeled as ``Gaussian square" in shape. For each set of parameters, CR gate amplitudes are tuned up in a two-pulse echo for a fixed gate time by minimizing the Bloch vector~\cite{QV64_2020}. A rotary pulse is additionally applied to the target qubit during the CR pulses, the amplitude of which is optimized to minimize gate error, as described in~\cite{Sundaresan2020}. Using an adaptive Runge-Kutta solver, we calculate the full unitary evolution of the system resulting from the calibrated gate sequence. From this simulated unitary $U$, error can be estimated by $E=1-d*(|trace(U_{\rm target} U^\dagger|^2+1)/(d+1)$ (where $d=2^n$ and $n=2$ for our two-qubit simulations) and is compared to the results of two-qubit randomized benchmarking.

\bibliographystyle{IEEEtran}
\bibliography{references}

\section*{Acknowledgments}
The authors acknowledge support for collision yield modeling and multi-qubit device characterization work under IARPA (Contract No. W911NF-16-0114). The authors also thank the IBM Microelectronics Research Laboratory and the Central Scientific Services for assistance with chip fabrication and handling, as well as associated control electronics. We thank M. Carroll, A. Rosenbluth and J. Gambetta for helpful discussions and feedback. We acknowledge the broader IBM team for their project support.

\section*{Data availability}
The experimental and numerical simulation data found in this manuscript are available from the corresponding author upon reasonable request.

\section*{Author contributions}
E.J.Z, Y.M. and J.S.O. constructed the LASIQ tuning system. E.J.Z., J.M.C. and J.S.O. conceived the experiments. S.S., N.S., D.F.B. and D.T.M. performed coherence and gate fidelity measurements of the multi-qubit chips. S.S., N.S., C.W., G.A.K., I.L., M.B.R., O.E.D. and M.B. helped design and fabricate the devices. J.B.H. built the Monte Carlo model for yield and collision simulations. E.J.Z. and J.B.Y. measured room temperature junction resistances. E.J.Z. performed the LASIQ tuning trials. J.S.O. and S.R. developed the LASIQ tuning method. J.T. and E.J.P. developed the numerical models for gate error estimation. W.L., E.P.L. and A.N. performed pre-cooldown preparations for the multi-qubit chips. E.J.Z., J.S.O., S.S., N.S., J.T., E.P. and J.M.C. prepared the manuscript with input from all authors.

\newpage
\section{Supplementary Information}
\label{sec:supplement}

\captionsetup[figure]{
  font      = footnotesize,
  labelfont = {bf},
  name      = {Supplementary},
  labelsep  = period,
  skip      = 12pt
}

\renewcommand{\thefigure}{S\arabic{figure}}
\setcounter{figure}{0}

In this supplementary section we present and expand upon results of (A) Monte Carlo yield modeling of the tuned (and comparison against the untuned) 27-qubit \emph{Falcon} processor presented in \hyperref[Sec II-A]{Sec II-A}, and (B) Statistical analysis of tuning success rates as described in \hyperref[Sec. II-B]{Sec.~II-B}.

\subsection{27-qubit Falcon yield}
\label{suppl:FalconYield}

In \hyperref[Sec II-A]{Sec II-A}, an example 27-qubit \emph{Falcon} processor was LASIQ-tuned for nearest-neighbor (NN) type 1-4 collision removal. To ascertain the impact of tuning on the NN collision-free yield rates, a Monte Carlo simulation \cite{Hertzberg2020} has been performed for the initial (pre-tuned) and final (post-tuned) transmon qubit frequencies, as predicted by typical power-law \emph{f\textsubscript{01}(R\textsubscript{n})} curves for our sample. As described in \cite{Hertzberg2020}, our model incorporates a random deviation (i.e. normal spread) from the target qubit frequencies, which may arise from empirical qubit spread from cleaning, bonding and cooldown processes (18.5~MHz), tuning imprecision (4.7~MHz), or other factors resulting in deviations from the desired target \hyperref[Sec. II-B]{(Sec.~II-B)}. We model the yield by adding to each target qubit frequency a random variation consistent with this frequency spread, and count the proportion of NN collision-free outcomes.

\begin{figure}[t!]
\label{suppl:1}
\centering
\includegraphics[width=7in, trim= {0.09in 3.9in 2.3in 0.11in}, clip]{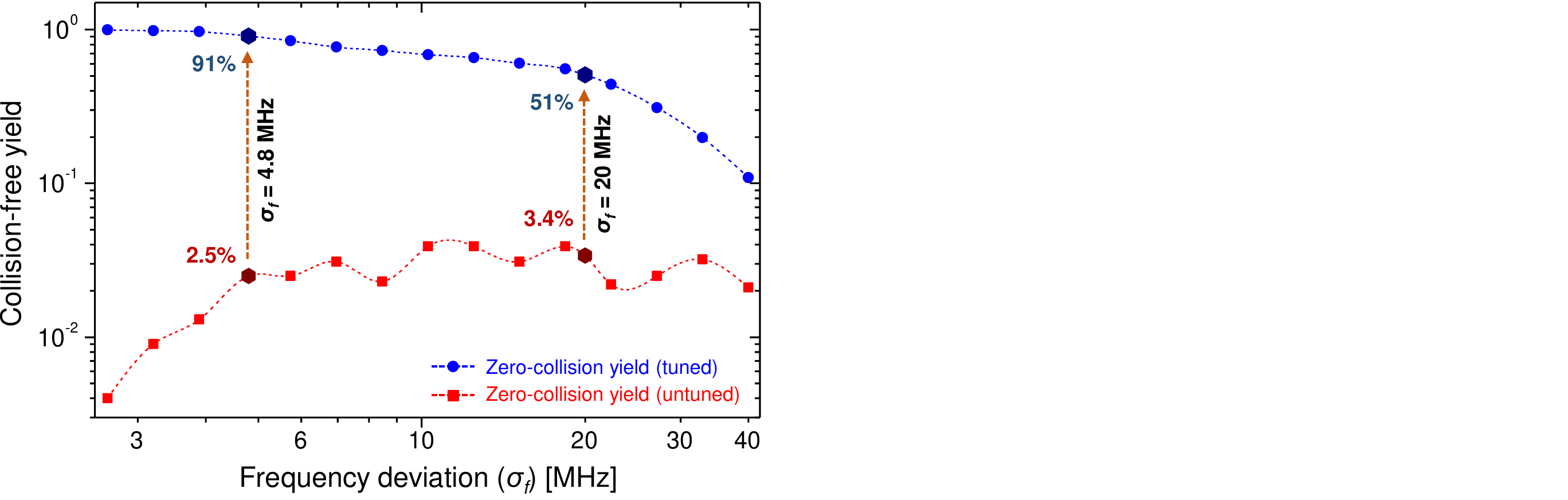}
\caption{Monte Carlo yield modeling for nearest-neighbor (NN) collision-free operation of a 27-qubit \emph{Falcon} processor, based on pre- and post-LASIQ frequency predictions determined in \hyperref[Sec II-A]{Sec II-A} (\hyperref[fig:1]{Fig. 1}). Pre-tuning (red), the yield curve is poorly conditioned and indicates $<$~5\% NN collision-free yield for frequency deviations up to 40~MHz. Low yields for small spreads are due to the presence of initial type 1-4 collisions that exist on the as-fabricated chip. Post-tuning, the final yield curve (blue) indicates $\geq$~10$\times$ improved yield rates. For nominal post-tuned spreading ($\sim$20~MHz, \hyperref[Sec. II-C]{Sec.~II-C}), our models indicate 51\% NN collision free yield. At the limit of LASIQ tuning precision predicted for this chip (4.8~MHz), this yield improves to above 90\%.
}
\end{figure}

\hyperref[suppl:1]{Fig. S1} shows the result of our yield models, indicating both the untuned (red) and tuned yield rates (blue) as a function of target frequency deviation. As-fabricated, the chip yields a significant number of type 1-4 NN collisions, as evidenced by the poorly conditioned untuned yield curve (red). The low yields for small \emph{$\sigma$\textsubscript{f}} results from the presence of existing collisions immediately after fabrication and prior to tuning. However, even accounting for large random spreads, the initial untuned lattice frequency pattern is highly collision prone, with yields remaining under 5\% for \emph{$\sigma$\textsubscript{f}}~$<$~40~MHz. The poorly conditioned untuned yield curve is a general characteristic of fixed-frequency architectures that exhibit frequency crowding. The primary goal of our LASIQ tuning method to arrange our lattice frequency into well conditioned yield curves with high probability of achieving a NN collision-free configuration.

\begin{figure}[t!]
\label{suppl:2}
\centering
\includegraphics[width=7in, trim= {-0.25in 4.7in 2.95in 0.15in}, clip]{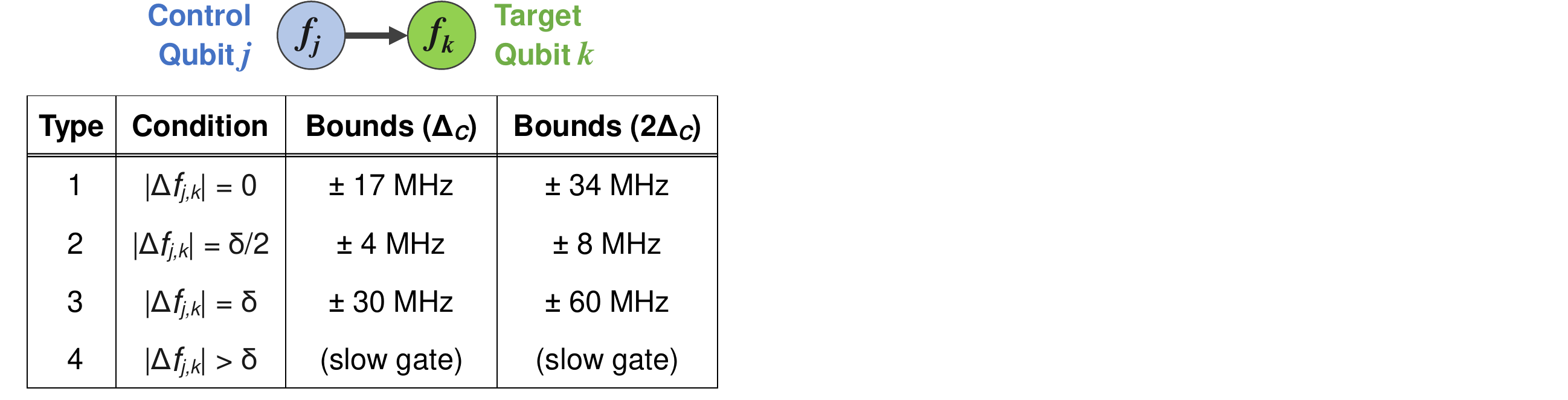}
\caption{Nearest-neighbor (NN) frequency collision bounds utilized in Monte Carlo yield modeling and the generation of NN collision-free tuning plans. Four collision types are identified, which correspond to level hybridization (type-1), excitation of the control or target qubit into the non-computational $\ket{2}$ state (type-2 and type-3), as well as `slow gates' (type-4) where excess two-qubit detuning results in diminished ZX interaction. Qubit anharmonicities are engineered near $\delta$~$\simeq$~\textendash330~MHz (see \hyperref[meth:FabricationAndPreparation]{Methods}). The quantitative collision bounds ($\Delta$\textsubscript{c}) corresponding to gate errors below 1\% are displayed in the third column, as determined in \cite{Magesan2020,Hertzberg2020}. The final column (2$\Delta$\textsubscript{c}) indicates double bounds utilized in generating NN collision-free tuning plans, serving as a `best practice' protocol for successfully yielding NN collision-free candidates.
}
\end{figure}

\hyperref[suppl:2]{Fig. S2} tabulates the various NN collision types, as well as tolerance bounds around which collision avoidance is successfully achieved. As described in the main text and elaborated in \cite{Hertzberg2020}, four main collision types are enumerated. Type-1 corresponds to two-qubit hybridization, with degenerate $\ket{0}$$\rightarrow$$\ket{1}$ control (\emph{j}) and target (\emph{k}) level spacings. Type-2 results from two-photon excitation of \emph{j} into the non-computational $\ket{2}$ state, and type-3 poses similar risk when $\emph{f\textsubscript{j,01}}=\emph{f\textsubscript{k,12}}$ and vice-versa. Type-4 corresponds to a `slow gate,' whereby reduced ZX interaction strength occurs due to two-qubit detuning beyond the straddling regime, resulting in longer gate operating times. The $\Delta\textsubscript{c}$ collision bounds are displayed in the third column of \hyperref[suppl:2]{Fig. S2}, numerically determined to yield $\leq$~1\% two-qubit gate errors \cite{Hertzberg2020,Magesan2020}, and are consequently utilized in the Monte Carlo collision and yield models depicted in \hyperref[suppl:1]{Fig. S1}. In the process of generating tuning plans for LASIQ, we generally adopt the `best practice' protocol of doubling these to 2$\Delta\textsubscript{c}$ bounds shown in the final column of \hyperref[suppl:2]{Fig. S2}, which provide tuning plans with greater robustness to post-tuned drifts, and therefore yields NN collision-free tuned candidates with increased confidence.

The tuning plan for this particular \emph{Falcon} processor is outlined in \hyperref[fig:1]{Fig. 1(a)}, with the aim of tuning out NN collisions and ensuring frequencies reside within the Purcell filter bandwidth, while maintaining resistance tuning $<$~14\%. All qubits are successfully tuned to frequency targets with a predicted 4.8~MHz precision, as outlined in \hyperref[Sec II-A]{Sec II-A}. The resulting Monte Carlo yield model of the tuned lattice pattern is shown by the well-conditioned (blue) curve in \hyperref[suppl:1]{Fig. S1}, indicating a typical yield roll-off as the frequency deviation increases. For nominal practical spreads ($\sim$20~MHz, \hyperref[Sec. II-B]{Sec.~II-B}), we calculate a yield improvement of 15$\times$, from 3.4\% to 51\%, indicating the efficacy of the LASIQ tuning method. A similar comparison is performed for the baseline precision achievable by LASIQ, which is predicted at 4.8~MHz for this chip (based on 0.16\% post-LASIQ resistance precision), and indicates NN collision-free yields beyond 90\%. As lattice sizes are scaled to 10\textsuperscript{3} qubit levels, practical tuning precisions approaching these single-MHz levels will be required to ensure appreciable yields. Presently, a major contributor to our practical trimming precision results from the imprecision of predicting \emph{f\textsubscript{01}} from \emph{R\textsubscript{n}} \cite{Hertzberg2020}, and significant gains may be achieved by leveraging processor recool cycles to tailor \emph{f\textsubscript{01}(R\textsubscript{n})} predictions. In particular, statistics on qubit frequency deviations during cryogenic cycling yields a recool stability of 5.7~MHz (see \hyperref[Discussion]{Sec. III}), which provides a practical pathway to approach the baseline precision of $\sim$5~MHz set by the frequency-equivalent resistance precision of our LASIQ process.

\subsection{Tuning success statistics}
\label{suppl:SuccessStats}

In \hyperref[Sec. II-B]{Sec.~II-B}, a tuning sample of 349 \emph{Falcon} qubits out of 390 total qubits were tuned to resistance targets (\emph{R\textsubscript{T}}), spanning a tuning range up to 14.5\% with 89.5\% success rate. In this section we break down the tuning success statistics of the tuned qubits, and outline failure mechanisms which induce deviations from target resistances extending beyond the desired target band.

\begin{figure}[t!]
\label{suppl:3}
\centering
\includegraphics[width=7in, trim= {0.13in 3.05in 5.2in 0.13in}, clip]{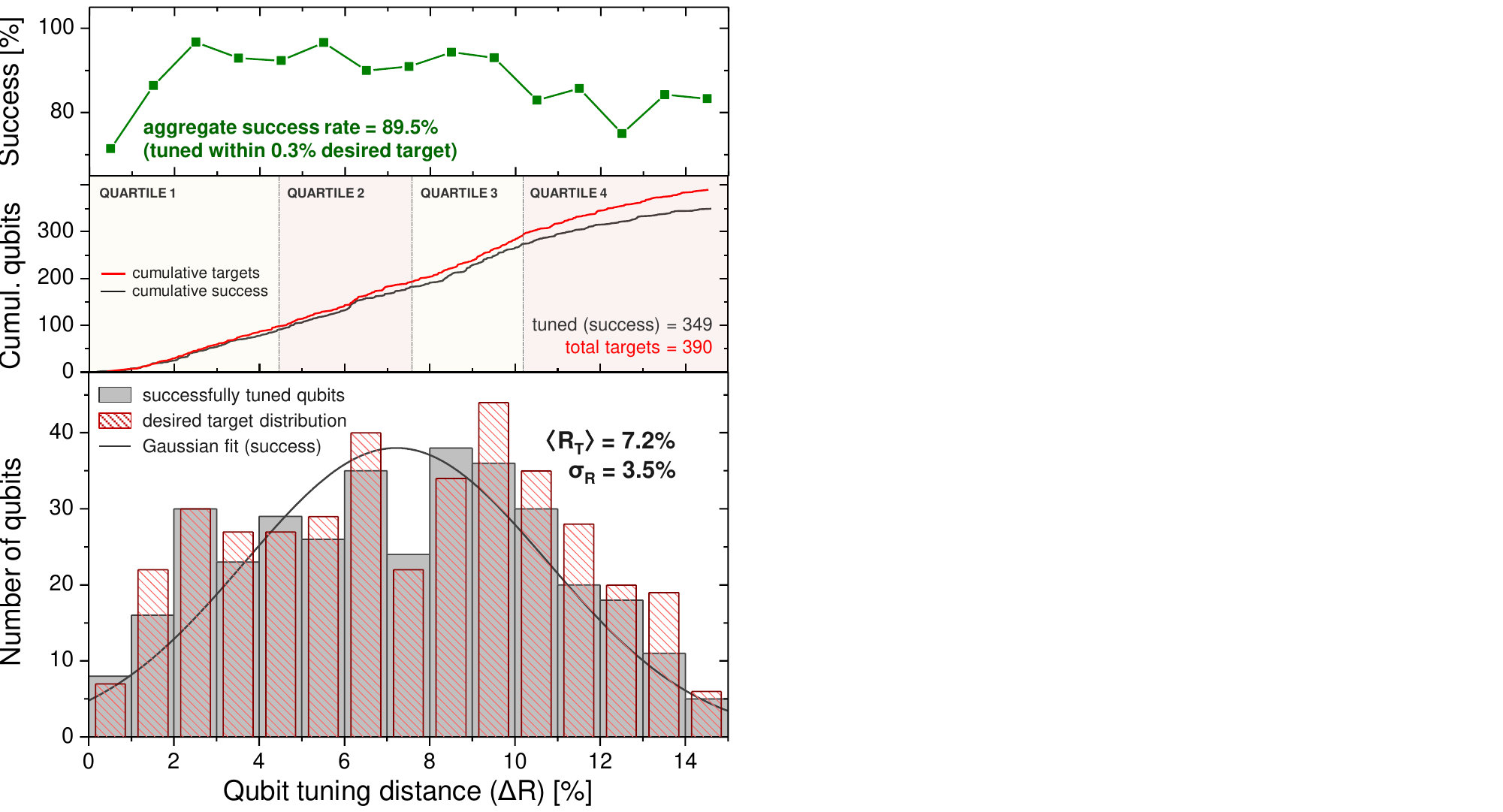}
\caption{Aggregate statistics of 390 qubits from trial tuning iterations of \emph{Falcon} processors. The bottom panel indicates the desired tuning resistance targets and actual tuning distance (\emph{$\Delta$R}, shown by the striped red and shaded gray histograms respectively). A Gaussian distribution over the successfully tuned qubits (349 out of 390) is shown, with a mean tuning range of 7.2~$\pm$~3.5\%. The cumulative number of tuned qubits is shown in the middle panel, with total tuned qubits (red) and successfully tuned qubits (black) displayed for comparison. The deviations demonstrate that at higher desired tuning distances, a greater failure rate occurs due to difficulty reaching the desired \emph{$\Delta$R}. This is further corroborated in the top panel (green squares) indicating the tuning success rate for each \emph{$\Delta$R} class of qubits. Near the wings of the curve, we observe dips in success rate, with the junctions with lower target \emph{$\Delta$R} suffering generally from overshoot, whereas higher target \emph{$\Delta$R} values suffer from undershoot. The total aggregate success rate (defined as tuning to within 0.3\% target resistance band) is 89.5\%.
}
\end{figure}

\hyperref[suppl:3]{Fig. S3} shows the tuning statistics of all 390 qubits used in our precision benchmarking experiments. The bottom panel indicates the desired tuning histogram distribution (red, striped) with respect to initial junction resistance, superimposed upon the successfully tuned qubits (gray). Tuning success is defined as tuning to within 0.3\% band around \emph{R\textsubscript{T}}, and the LASIQ system will proceed incrementally towards this goal. A Gaussian distribution of the successfully tuned qubits is also depicted, with a mean tuned distance of 7.2\% ($\sigma$~=~3.5\%). We note here that each of the \emph{Falcon} qubits was tuned as part of a tuning plan designed for NN collision avoidance, and hence our tuning distances are representative of realistic processor samples. It is notable that deviations between the desired (red, striped) and actual (gray) tuning distributions increasingly depart at both lower and higher tuning ranges, indicating that missed targets occur at the tuning distribution extremities.

The central panel of \hyperref[suppl:3]{Fig. S3} indicates the cumulative plot of LASIQ tuned qubits, with both tuned total (red) and successfully tuned (black) qubits shown. Corresponding success rates for each tuning distance (assessed at 1\% intervals) are displayed in the top panel, which indicates that failure rates are most likely ($<$~90\% success) for tuning targets under 1\% and above 10\%. Below, we consider each case in turn.

In the former case, small desired tuning distances are exposed to risk of overshooting (i.e. \emph{$\Delta$R~$>$~$R\textsubscript{T}-R\textsubscript{n}$}). This is a result of the statistical spread in resistance progression, and small desired resistance shifts suffer increased risk of accidental increase beyond desired targets. Despite this risk, we note that typical optimal frequency separation between qubits in our tuning plans are on the order $\sim$100~MHz, and even for significant overshoot, NN collisions are unlikely to be induced unless the deviation between tuned resistances and targets approach the frequency spacing between NN qubits.

In the latter case, undershoot effects appear for junction targets beyond \emph{$\Delta$R}~$>$~10\%, as evidenced by the slow tapering down of success rates in the top panel of \hyperref[suppl:3]{Fig. S3}. This is a consequence of junctions reaching their maximum tuning limit, which based on calibration measurements on junction arrays, is near 14\%, with a spread of $\sigma$~$\sim$~2\%. In almost all cases, tuning plans are engineered to reside under 14\% tuning, to ensure that almost all qubits on a given processor have high probability of reaching desired targets. Nevertheless, that statistical nature of the tuning range causes outlier qubits which saturate at lower \emph{$\Delta$R}, resulting in monotonically decreasing success rates as observed in \hyperref[suppl:3]{Fig. S3}. Progress in scaling lattices of transmon qubits will benefit greatly from determining the origin of this resistance limit, and the best-practice fabrication processes through which this may be increased.

\vspace{0.20in}
\hspace{0.05in}
\rule{3.0in}{0.5pt}
\end{document}